# A Research Review on Detection and Classification of Power Quality Disturbances caused by Integration of Renewable Energy Sources

Utkarsh Singh[*]

September 24, 2020

**Abstract-** With the increased interest in integrating renewable energy sources (RES) such as wind power and solar into the power systems owing to their zero greenhouse gas emissions and the involvement of power converters for integration in grid, the detection, classification and mitigation of power quality events has become indispensable. For employing an appropriate mitigation technique, it is a pre-requisite to correctly classify the various types of disturbances in power quality. This paper, therefore, presents a detailed research reviews on detection and classification of power quality disturbances caused by the integration of renewable energy sources and associated works present in literature till date. Attempts are also made to highlight the current and future issues involved in the detection, classification and mitigation of PQ disturbances. Best efforts have been made to make this paper serve as a full-fledged reference for the future work in this field. A list of 230 research publications on the subject is also appended for quick reference.

## 1. Introduction

With the increased concern over global climate change and depletion of fossil fuels, utilization of renewable energy sources such as wind power and solar has grown in the past decades, annually. Integration of these energy sources into the grid has created complexity in the operation of power systems due to their intermittent nature. The major impacts of wind and solar integration are unbalance and voltage fluctuations, voltage and current harmonics, grid islanding protection, and other power quality issues, such as flicker and stress on distribution transformer. Severity of these issues depends on the penetration level of these energy sources, their location and configuration of distribution system.

Before moving on to power quality issues, one should have a clear understanding of 'Power Quality'. One of the earliest references [1], defines power quality as a tool for describing the conditions at the interface of power sources and loads. To avoid misinterpretation of the term power quality, five different terms were defined [2]: voltage quality, current quality, power quality, quality of supply and quality of consumption. For ease of understanding, it may be said that power quality deals with the deviations in voltage and current waveforms from their respective ideal behaviors in a power system, from the point of generation till consumption. These deviations in voltage or current waveforms may be termed as power quality disturbances.

---

[*] The author is with Artificial Intelligence Lab, Free University of Brussels (VUB), 1050 Brussels, Belgium. Corresponding e-mail: utkarsh.singh@ieee.org

Various types of PQ disturbances or events have been discussed in literature and many techniques for their detection, localization and classification have also been presented. The methods differ in complexity, hardware requirements, computational speed, cost of implementation, suitability and popularity. These techniques range from most innovative ideas (but not effective) to simple methods (yet effective). Due to abundance of such techniques in literature, it is very necessary to categorize which method should be adopted for a particular problem. A survey would be beneficial here, for the researchers in power quality. Fig. 1 gives an approximate research trend ranging from the earliest traceable work to till date. It can be observed from the trend that the interest in power quality has declined in last five years. With the help of this review, it will be shown that there is much more to be uncovered in power quality and classification of disturbances, especially with the emergence of smart grids, micro grids and renewable energy.

Over 200 papers related to PQ and disturbance classification have been compiled in this manuscript. It has not been intended to present a literal chronology of all the work done in this context, because the publication date is generally not indicative of when a particular technique was adopted. It has also been tried to omit papers with reference to a previous work without any significant modification. Author apologizes if one or more important works have been omitted unintentionally. This paper, thus presents a state-of-the-art discussion on PQ analysis and disturbance classification, and has been divided into following segments- Power quality overview, Signal processing techniques, detection and localization, classification techniques and conclusion.

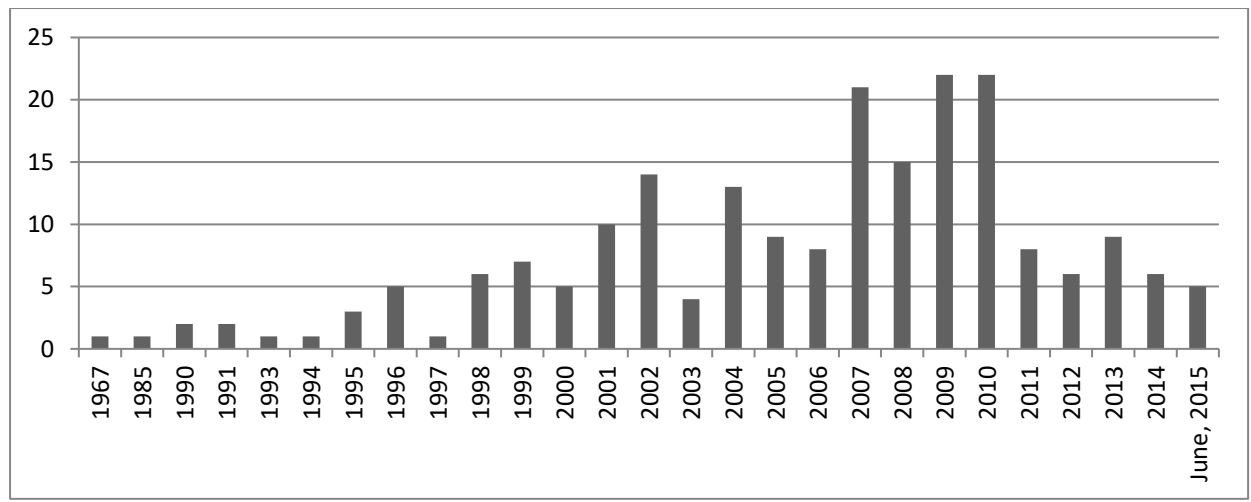

Fig. 1. Total number of papers per year, since 1967

## 2. Power quality overview

This section will focus on various aspects of power quality and PQ events. The definition of power quality must not be limited to source-load interactions as it also depends on the interaction between equipment and electromagnetic environment. This interaction is termed as electromagnetic compatibility (EMC) and the international power quality standards (IEC) are considered as a subset of EMC. Depending on how the power quality disturbances are measured, these can be broadly classified into two types: Variations and Events. Variations are the small deviations (rms or frequency) in voltage or current from their nominal value, e.g. harmonic distortion, voltage fluctuation, voltage and current unbalance and high frequency voltage noise. Variations are measures at a particular time. Events are larger deviations, which occur occasionally e.g. load switching currents or momentary interrupts. Events are measured w.r.t start and end time and deviations crossing a threshold value, e.g. voltage sag, swell interruptions and transients [2]. Power quality disturbances (PQD) may arise due to variations in amplitude, frequency and waveform. Depending upon the duration of existence, events may be classified as short, medium or long type [3]. Different types of classification schemes may be found in literature, but a simplified classification may be divided into the following types [4]: Interruption, under voltage, over voltage, voltage/current unbalance, harmonics (integer harmonics, inter harmonics and sub harmonics), transients (oscillatory and impulsive), voltage sag, voltage swell, flicker, notch and noise. In order to classify a disturbance, it is very necessary to properly understand the pattern of a PQD [5], therefore the nature of commonly used PQDs has been listed in Table 1 [3,4]. The common sources of these disturbances are power electronic devices, IT and office equipments, arcing devices, load switching, large motor starting, embedded generation, sensitive equipments and environment related damage [6]. The equations for modeling the commonly classified events have been given in Table 2, and may also be found in previous works [7-15]. The literature on power quality is increasing but it will not saturate as a research topic due to power system modernization. For detailed knowledge of power quality, readers are encouraged to go through the publications listed in references [16-43].

Table 1: Power quality disturbances and their description [3,4]

| Disturbance | Description |
|---|---|
| Voltage sag | Reduction in RMS voltage over a range of 0.1–0.9 pu for a duration greater than 10 ms but less than 1 s. |
| Voltage swell | increase in RMS voltage over a range of 1.1–1.8 pu for a duration greater than 10 ms but less than 1 s. |
| Under voltage | Voltage magnitude is below its nominal value. |
| Over voltage | Voltage magnitude is above its nominal value. |
| Interruption | Voltage magnitude is zero. |
| Flicker | A visual effect of frequency variation of voltage in a system. |
| Voltage/current unbalance | Deviation in magnitude of voltage/current of any one or two of the three phases. |
| Outage | Power interruption not exceeding 60 s duration due to fault or mal-tripping of switchgear/system. |
| Transients | Sudden rise of signal which may be impulsive or oscillatory in nature. |

| Harmonics | Non-sinusoidal waveforms having frequency other than fundamental frequency. |
|---|---|
| Integer harmonics | Harmonics with frequency as integer multiple of fundamental frequency. |
| Inter harmonics | Harmonics with frequency higher than fundamental frequency, but not integer multiples of it. |
| Sub harmonics | Harmonics with frequency lower than fundamental frequency. |
| Notch | Non-sinusoidal, periodic waveform distortions. |
| Noise | Low magnitude electrical signals from a broad frequency spectrum lower than 200 kHz. |

Table 2: Models of PQ events

| Disturbance | Equation | Parameters |
|---|---|---|
| Normal | $x(t) = \sin(\omega t)$ | $\omega = 2\pi \cdot 50 \ rad/s$ |
| Sag | $x(t) = [1 - \alpha(u(t - t_1) - u(t - t_2))] \sin(\omega t)$ | $0.1 \leq \alpha \leq 0.9, T \leq t_2 - t_1 \leq 9T$ |
| Swell | $x(t) = [1 + \alpha(u(t - t_1) - u(t - t_2))] \sin(\omega t)$ | $0.1 \leq \alpha \leq 0.8, T \leq t_2 - t_1 \leq 9T$ |
| Interruption | $x(t) = [1 - \alpha(u(t - t_1) - u(t - t_2))] \sin(\omega t)$ | $0.9 < \alpha \leq 1, T \leq t_2 - t_1 \leq 9T$ |
| Flicker | $x(t) = [1 - \alpha \sin(2\pi\beta t)]\sin(\omega t)$ | $0.1 \leq \alpha \leq 0.2, 5Hz \leq \beta \leq 20Hz$ |
| Oscillatory transient | $x(t) = \sin(\omega t) + \alpha \exp(-\frac{t-t_1}{\tau})(u(t - t_1) - u(t - t_2)) \sin(2\pi f_n t)$ | $0.1 \leq \alpha \leq 0.8$, $0.5T \leq t_2 - t_1 \leq 3T$, $300Hz \leq f_n \leq 900Hz$, $8ms \leq \tau \leq 40ms$ |
| Harmonic | $x(t) = \alpha_1 \sin(\omega t) + \alpha_3 \sin(3\omega t) + \alpha_5 \sin(5\omega t) + \alpha_7 \sin(7\omega t)$ | $0.05 \leq \alpha_3, \alpha_5, \alpha_7 \leq 0.15, \sum(\alpha_i)^2 = 1$ |
| Notch | $x(t) = \sin(\omega t) - sign(\sin(\omega t))\{\sum_{n=0}^{9} \kappa * [u(t - (t_1 + 0.02n)) - u(t - (t_2 + 0.02n))]\}$ | $0.1 \leq \kappa \leq 0.4, 0 \leq t_1, t_2 \leq 0.5T, 0.01T \leq t_2 - t_1 \leq 0.05T$ |
| Spike | $x(t) = \sin(\omega t) + sign(\sin(\omega t))\{\sum_{n=0}^{9} \kappa * [u(t - (t_1 + 0.02n)) - u(t - (t_2 + 0.02n))]\}$ | $0.1 \leq \kappa \leq 0.4, 0 \leq t_1, t_2 \leq 0.5T, 0.01T \leq t_2 - t_1 \leq 0.05T$ |
| Sag with harmonic | $x(t) = [1 - \alpha(u(t - t_1) - u(t - t_2))] * [\alpha_1 \sin(\omega t) + \alpha_3 \sin(3\omega t) + \alpha_5 \sin(5\omega t)]$ | $0.1 \leq \alpha \leq 0.9$, $T \leq t_2 - t_1 \leq 9T$, $0.05 \leq \alpha_3, \alpha_5 \leq 0.15$, $\sum(\alpha_i)^2 = 1$ |
| Swell with harmonic | $x(t) = [1 + \alpha(u(t - t_1) - u(t - t_2))] * [\alpha_1 \sin(\omega t) + \alpha_3 \sin(3\omega t) + \alpha_5 \sin(5\omega t)]$ | $0.1 \leq \alpha \leq 0.8$, $T \leq t_2 - t_1 \leq 9T$, $0.05 \leq \alpha_3, \alpha_5 \leq 0.15$, $\sum(\alpha_i)^2 = 1$ |

## 3. Power quality indices (PQI)

Certain indices (known as power quality indices) have been prepared to minimize the dissimilarity of results while monitoring power quality using different techniques. The power quality indices (PQI) include parameters such as rms, magnitude, phase, frequency or energy of signals. The deviation in these indices during a disturbance as compared to their nominal or threshold values may be then used in accordance with the standards [38,44] for detection or categorization of PQDs. Other standards for measurement of parameters for detection of power quality events have been mentioned in [35,45,46]. Mathematical representations of these PQI can be found in [47].

## 4. Detection

Power quality indices or features of a power signal are used for detection of power quality disturbances or events. Signal segmentation is a necessity for reduced data handling and it also helps in categorizing the events as pre- and post-fault events. Wavelet transform can be used along with threshold check for detection and time estimation of disturbances [48-52]. Use of adaptive linear combiner (ADALINE) with wavelet transform may help in residual detection and reduce the burden of training [53,54]. Authors have emphasized on the necessity of test sample size and calculation time reduction for real time power quality analysis [55,56]. Fundamental voltage component provides high precision level and works well even in presence of DC decaying components. Peak voltage does not provide much detail about PQD but is a cheaper option from detection perspective. RMS voltage is suitable for harmonic and flicker detection. Fundamental voltage component is good for detection of sag, swell and interrupts. Peak voltage based detection has been suggested for impulses and surges [57]. It was found that one window length is suitable for short duration disturbance detection. Time at level analysis was suggested as an improvement to the conventional RMS based detection techniques where RMS estimation was done using a moving window. Time at level curves is suitable for comparison with standard curves like CBEMA and ITIC [58]. Preset threshold advantage was presented in an easy VI program [59]. RMS per unit, waveform distortion, harmonic and interharmonic distortions are suitable indices for PQD detection using rule based systems [60]. Higher order cumulants and Phase locked loop (PLL) are good for transient detection and differentiating between long and short transients [61-63]. Spectral sincfit algorithm was given as a robust method for harmonic detection [64]. Generalized likelihood ratio test (GLRT) detector shows better performance in sag detection as compared to RMS, wavelet, kalman filtering, peak voltage and missing voltage based detection techniques [65].Various methods have been discussed in literature for sag/swell and rms based detection [56-58,60,66-68]. Effect of noise and harmonics on detection has also been discussed in some of these cases. Some other variants with simplified approach were: Fuzzy expert system [69], Space vector representation [70], neural network based detection [56,71,72], Phase Space Embedding [73], 2D wavelet transform [74], Phase space representation [75], Continuous wavelet transform [76], Adaptive filtering [77], Symmetrical components [78], Multiresolution morphological gradients

[79], Sparse signal decomposition [80] and Hilbert-Huang transform [81]. Phase space representation based techniques were found to perform better than window based techniques.

## 5. Signal processing techniques

Signal processing is the most vital part of PQD classification. It is very important to analyze the time and frequency components of a power signal to accurately detect and classify the events. Many techniques evolved in the course of time, out of which Wavelet transform and Stockwell transform gained huge popularity. Though these techniques can be event specific such as: parametric (fixed number of parameters) and non-parametric (parameters grow with the amount of training data) or according to the signals: stationary (frequency content does not change over time) and non-stationary (frequency content changes over time), but it is better to list the pros and cons of these methods so that the researchers can choose a technique accordingly.

Important techniques available in literature include: Short Time Fourier Transform (STFT) [82-84], Wavelet Transform (WT) [85-117], Stockwell Transform (ST) [118-135], Hilbert/Hilbert-Huang Transform (HT/HHT) [81,136-141], Gabor/Gabor-Wigner Transform (GT/GWT) [142-145], Cohen's class [146-151], Filters [46,152-157], Prony Analysis (PA) [158-163], Phase Space Embedding [73], Phase Locked Loop [63,164,165], Parametric Methods [166-171]. Application of some of these techniques for disturbance detection has already been discussed in the previous section. The description and mathematical definitions of these techniques have been properly tabulated in [172]. To avoid any inconvenience to the readers, the key findings of the references used in this segment may be summarized as follows:

- Fourier transform is a powerful tool for frequency analysis but is incapable of detecting the sudden changes. STFT solves this problem by using a small window, so that these types of disturbances can be considered as stationary during analysis.
- STFT, GT, Filter banks, WT, ST, PA, Cohen's class and parametric methods are well suited for non-stationary signal analysis.
- STFT is simple in implementation but has limited time-frequency resolution.
- DWT offers simultaneous time-frequency analysis. Multi Resolution Analysis (MRA) makes it possible to break a signal into continuous high-low frequency components. Performance of this technique depends on the choice of mother wavelet and the level of decomposition. Debauchies-4 (Db4) wavelet has been found to be most suitable for PQD classification. Wavelet transform is affected by the presence of noise and has cross-term problem. High decomposition levels increases the computational burden.
- ST is capable of localizing the real and imaginary components of spectrum in time. It is a phase corrected form of continuous wavelet transform with a Gaussian window. Cross-term problem can be avoided with this method. This technique is not suitable for harmonics. Use for real time applications is also not recommended, due to high computational complexity and execution time.

- HHT is helpful in obtaining and preserving the instantaneous frequency data, and hence it is suitable for real time applications. Both non-stationary and non-linear signals can be analyzed using this technique. It is limited in distinguishing different components in narrow signals.
- GT is a special case of STFT. It offers good time-frequency resolution and signal to noise ratio (SNR). It has ability to zoom into the segment of interest in a signal. Computational complexity in this technique is proportional to sampling frequency and it is not suitable for analyzing high frequency events. Addition of Wigner's distribution function may help further improvement of time-frequency resolution but window width plays a very critical role in this case.
- Cohen's class offers different tools for achieving high time-frequency resolution (Spectrogram, Wigner-Ville distribution, Choi-Williams distribution and reduced interference distribution). Proper kernel selection may reduce the interference problem.
- Filter banks help in frequency sub-banding and computational complexity is lesser. Inaccurate harmonic prediction and frequency band overlapping are common problems with this technique.
- Kalman filter (KF) is suitable for real time harmonic and transient detection. It also offers good SNR. With Extended Kalman Filter (EKF), changes in the parameters of distorted signal can be detected. It is not capable of simultaneous time and frequency decomposition.
- PA does not require frequency information before filtering and additional frequency estimators. It is suitable for analysis of transients, harmonics and oscillations. Slight mismatch in model may result in incorrect estimation of an event.
- Phase space embedding represents the parameters of a dynamic system in phase space. It is suitable for real time application as it requires current sample and a quarter cycle ago sample. It also provides noise immunity to a great extent.
- PLL provides accurate measurement of phase and frequency. It is also capable of synchronizing with the input signal. Harmonic and interharmonic estimation is inconvenient with technique.
- Parametric methods such as MUSIC, Yule-Walker and Auto-Regressive (AR) models give outstanding time-frequency resolution. Though these methods maintain essential signal information but require large model order and sufficient time for accurate prediction. Proper assumption of statistical distribution of signal is a pre-requisite.

## 6. Feature extraction

In earlier reviews, feature extraction has not been discussed in particular. So this segment has been dedicated to feature extraction. A feature, in reference to power signals, can be defined as the distinctive attribute of a power signal which later assists in the classification of power quality events. A feature vector can hence be defined as a set of such features. Feature extraction refers to the process of selection of important features from a signal which can be used for classification. The foundation and applications of feature extraction have been elucidated in [173]. In [74], 60 features were generated using number, location and amplitude of local maximum points in 2D wavelet transform spectrum. Energy extracted from coefficients of multi resolution analysis has been used as a prime feature in wavelet transform based works [89,103,113,174,181]. Binary features can be developed in wavelet transform by using five different types of indices: duration index, interruption index, flat index, rising index and falling index [100]. RMS and THD based features are very helpful in harmonic detection and classification [104]. In [131], five different types of energy measures have been shown. Some other papers on wavelet transform present the additional set of features for distinguishing PQ events, such as: fundamental component, total harmonic distortion (THD), number of peaks in wavelet coefficients, phase angle shift, oscillation number of missing voltage, lower harmonic distortion and oscillation number of rms variations [175,179]; time duration and peak based features [178]; mean, standard deviation, skewness, kurtosis, RMS, form factor, crest factor and fast fourier transform based features [181]; number of samples in a particular range of amplitude or beyond that range [183]. Magnitude and argument coefficients obtained from continuous wavelet transform have been used as features to extract desired band of transient signal for disturbance detection [111,182]. In [184], 64 feature vectors were developed using mean and standard deviation out of which 8 optimal feature vectors were identified based on principal component analysis. Amplitude factor, maximum/minimum values of signal, standard deviation of magnitude and phase contour are suitable features to be extracted from Stockwell transform (ST) [125,176]. If the coefficients of a transform possess sufficient information for distinguishing between one or more events, they can be directly used to form a feature vector as shown in [136,184]. Parallel feature extraction in time and frequency domain speeds up the computational process significantly [185]. While feature extraction, it should be noted that features should not be correlated and should be able to distinguish between various PQDs. If the feature values are not uniform, data must be normalized. Various methods for feature vector normalization may be found in [186]. Irrelevant and redundant features must always be rejected which do not provide any useful information. Hence, optimal selection of features based on their relevance is very necessary. Techniques for optimal feature selection have been discussed in [187-189].

## 7. Classification

To avoid the ill effects of PQDs on power system and the equipments within, mitigation is required. Mitigation measures are taken according to the type of disturbance. This calls for the classification of PQDs. In this segment various classification methods available in literature will be discussed and the important aspects will be summarized. Bayesian classifiers use maximum likelihood (ML) criteria for PQD classification along with probability density functions of features [82,190-193]. In [190], a rule based approach has been presented for time characterized classification and wavelet transform (WT) - hidden markov model (HMM) for frequency characterized classification. Nearest neighbor (NN) classifiers have been covered in [82,174,194,195]. In [174], NN based pattern recognition has been shown for online application. Artificial neural network (ANN) has already been proved in past for their optimization, pattern recognition and data clustering capabilities. Use of ANN for PQD classification has been shown in

[8,9,11,82,124,192,133,136,196-211]. Common types of ANN, which have been widely implemented for classification are: Back propagation neural network (BPNN), Multi-layer perceptron (MLP) and Radial basis function neural network (RBFNN). BPNN is suitable for training multi layered feed forward networks. MLP offers good recognition ability. However, a proper scheme or optimization is required for proper selection of number of hidden layers and nodes. While MLP is slow, RBFNN offers fast learning and detection advantages. The membership of a particular PQ event to a class may be characterized by using fuzzy classifier [82,103,165,175,207,210,212-214]. The performance can be further improved by using neuro-fuzzy [175,212] or genetic-fuzzy combinations [207]. These hybrid schemes are used to improve the rule base of fuzzy classifier. Some rule based expert systems (ES) have been presented in literature [82,157,206,207,209,215-218], out of which fuzzy-expert system is most successful. Rule based decision trees are used in ES. A binary feature matrix scheme for classification has been presented in [215]. Support vector machine (SVM) is a very useful technique for auto-classification of disturbances. Use of SVM and its variants like multi-class SVM, fuzzy-SVM, directed acrylic graph SVM has been shown in [15,170,202,217,219-225]. Dynamic time warping classifier has been presented in [226], which is suitable for automated monitoring and provides superior speed and accuracy as compared to neural network and fuzzy classifiers. Important findings to assess the performance and abilities of these classifiers have been summarized as follows:

- Bayesian classifiers are suitable for functions with Gaussian probability density, which should be known beforehand. Large computational cost is a disadvantage in this case.
- Nearest neighbor classifiers are very accurate in classifying mixed event problems. Noise hinders the performance of these classifiers.
- ANN is appropriate for real time classification, but the performance depends on the network architecture. Noise reduction techniques must be used to avoid any effect on accuracy.
- Fuzzy logic classifiers offer easy modeling but an additional training set is required for a new PQ event.
- Expert systems do not rely much on the amount of input data available. These classifiers are slow and costly. Moreover, the classification may be affected in a case where a particular event does not match any of the rules.
- SVM classifiers have high learning capability. It can handle large number of features and is suitable for quadratic optimization problems. Good classification accuracy can be achieved only with proper training.

## 8. Conclusion

This paper provides an absolute review for the classification of power quality disturbances (PQD). All the important terms related to power quality and PQD classification have been properly defined. The role of power quality indices has been focused upon to obtain comparable performance with different detection and classification tools. An extended literature has also been provided for the readers to understand the power quality standards. This paper covers almost all the important aspects of detection, signal processing, feature extraction and classification with respect to power quality problems. Mathematical models of commonly investigated power quality events have also been presented in this paper for the ease of use. Feature extraction plays a key role in PQD classification; therefore, a dedicated segment has been presented in this paper, which was absent in earlier reviews. Summarized points in signal processing techniques and classification segments will help the researchers to easily assess the pros and cons of a technique without going through extensive literature.

From this review, it is clear that the main causes of poor quality are dips, surges, transients and momentary interrupts. Early detection and characterization schemes offered by power quality monitoring standards IEEE 61000-4-30 and IEEE 1159 were based on RMS voltage. The performance of these methods depends on window length, which is not suitable for transients. Several other techniques are popular for PQI estimation such as WT and filter banks. But there is a need for new set of standards keeping in view the estimation and classification of transients and harmonics. Singular point (start and end points of a disturbance) detection has recently emerged as a new area of interest. The conventional singular point detection schemes such as High Pass Filter and RMS method introduce time delay, are susceptible to noise and do not work well under frequency changes. With the advent of near perfect reconstruction (NPR) filter banks and improvement in design of digital filters, the accuracy of harmonic measurement has significantly improved. Among the signal processing techniques WT and ST have been used extensively, but these methods have the respective disadvantages of noise susceptibility and large computational burden, which makes them unsuitable for real time operation. Other techniques, which offer good time-frequency resolution are either very costly or fail to analyze the harmonic and transients disturbances properly. Rule based expert systems, fuzzy classifiers, artificial neural network (ANN) and support vector machines (SVM) are the common classifiers based on artificial intelligence. The SVM has become a preferred choice over the ANN recently. The main advantages of SVM are simple geometric interpretation and existence of global minima in all cases. In context of modern grids (smart and micro), it can be suggested that there is a need to monitor the nature of PQ events on these sites and prepare detection/classification schemes accordingly. A proper scheme is still required which offers speed, accuracy, uses minimum resources, less training and is adaptable to new disturbances. The various stages involved such as detection, signal processing, feature extraction and classification already increase the computational burden. Therefore, optimal feature selection is also indispensable to reduce the memory usage and computational time.

Substantial progress has been made in the area of PQ disturbances covering analysis, simulation, and hardware development and testing for identification, classification and mitigation etc. However, many problems and issues, especially those related to development of real time automated detection, classification and mitigation schemes etc., still need to be addressed for appropriate system planning and operation of power system to supply a good quality and reliable electric power.